\documentclass{vldb}
\usepackage{graphicx}
\usepackage{balance}  
\usepackage{booktabs} 
\usepackage{datadog}
\usepackage{multirow}
\usepackage{url}

\vldbTitle{DDSketch: A fast and fully-mergeable quantile sketch with
  relative-error guarantees}
\vldbAuthors{Charles Masson and Jee E Rim and Homin K. Lee}
\vldbDOI{https://doi.org/10.14778/3352063.3352135}
\vldbVolume{12}
\vldbNumber{12}
\vldbYear{2019}
\vldbPages{2195-2205}

\begin{document}

\title{DDSketch: A Fast and Fully-Mergeable Quantile Sketch with Relative-Error
  Guarantees}

\author{
\numberofauthors{3}
\alignauthor
Charles Masson\\
       \affaddr{Datadog}\\
       \affaddr{620 8th Ave.}\\
       \affaddr{New York, NY}\\
       \affaddr{charles.masson@datadoghq.com}
\alignauthor
Jee E. Rim\\
       \affaddr{Datadog}\\
       \affaddr{620 8th Ave.}\\
       \affaddr{New York, NY}\\
       \affaddr{jee.rim@datadoghq.com}
\alignauthor
Homin K. Lee\\
       \affaddr{Datadog}\\
       \affaddr{620 8th Ave.}\\
       \affaddr{New York, NY}\\
       \affaddr{homin@datadoghq.com}
}

\maketitle

\begin{abstract}
  Summary statistics such as the mean and variance are easily maintained for
  large, distributed data streams, but order statistics (i.e., sample quantiles)
  can only be approximately summarized. There is extensive literature on
  maintaining quantile sketches where the emphasis has been on bounding the rank
  error of the sketch while using little memory. Unfortunately, rank error
  guarantees do not preclude arbitrarily large relative errors, and this often
  occurs in practice when the data is heavily skewed.

  Given the distributed nature of contemporary large-scale systems, another
  crucial property for quantile sketches is mergeablility, i.e., several
  combined sketches must be as accurate as a single sketch of the same data.
  We present the first fully-mergeable, relative-error quantile sketching
  algorithm with formal guarantees. The sketch is extremely fast and accurate,
  and is currently being used by Datadog at a wide-scale.
\end{abstract}

\section{Introduction}

Computing has increasingly moved to a distributed, containerized, micro-service
model. Some organizations run thousands of hosts, across several data centers,
with each host running a dozen containers each, and these containers might only
live for a couple hours~\cite{Datadog-docker,Datadog-orchestration}. Effectively
being able to administer and operationalize such a large and disparate fleet of
machines requires the ability to monitor, in near real-time, data streams coming
from multiple, possibly transitory, sources~\cite{BJPM-2016}.

The data streams that are being monitored can include application logs, IoT
sensor readings~\cite{MFHH-2003}, IP-network traffic
information~\cite{CJSS-2003}, financial data, distributed application traces~\cite{SFSG-2014},
usage and performance metrics~\cite{AABBCGMMRSWZ-2013}, along with a myriad of other measurements and
events. The volume of monitoring data being transmitted to a central processing
system (usually backed by a time-series database or an event storage system) can
be high enough that simply forwarding all this information can strain the
capacities (network, memory, CPU) of the monitored resources. Ideally a
monitoring system helps one discover and diagnose issues in distributed
systems---not cause them.

\begin{figure}
  \centering \includegraphics[width=.5\textwidth]{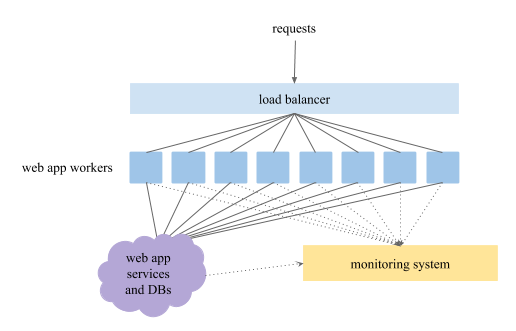}
  \caption{A distributed web application, with each container sending metrics to
    the monitoring system.}
  \label{fig:monitoring}
\end{figure}

\begin{figure}[hb]
  \centering
  \includegraphics[width=\columnwidth]{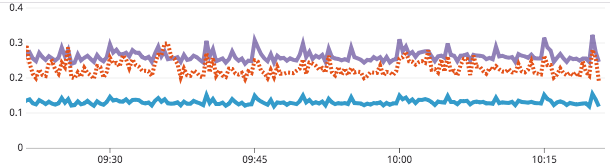}
  \caption{The average latency of a web endpoint over time (dotted line) is
    closer to the 75th percentile than it is to the 50th (the two solid lines).}
  \label{fig:p50meanp75}
\end{figure}

\begin{figure*}[!ht]
  \centering
  \includegraphics[width=0.8\textwidth]{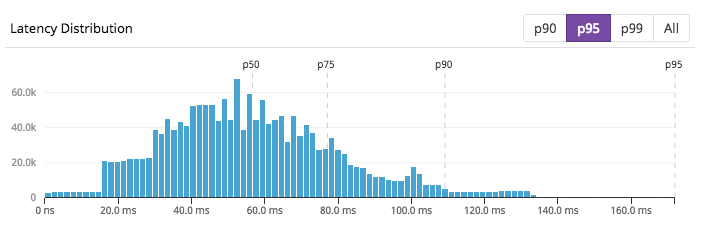}
  \includegraphics[width=0.8\textwidth]{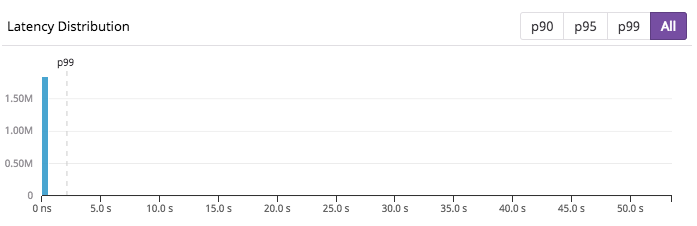}
  \caption{Histograms for p0-p95 and p0-p100 of 2 million web request response times.
    (Bars for p93-p100 exist but are shorter than the minimum pixel height.)}
  \label{fig:histograms}
\end{figure*}

Our running example will be a web application backed by a distributed system,
where the ability to answer any particular request might depend on several
underlying services and databases (Figure~\ref{fig:monitoring}). The metric
we monitor for our example will be the latency of the requests it handles. Every
time a worker finishes handling a request, it will note how long it took. Simple
summary statistics such as the overall mean and variance can be easily
maintained. For instance, the workers can keep counts, sums, and sums of squares
of the latency and send those values to the monitoring system (and reset those
values) every second. The monitoring system will then be able to aggregate those
values and derive metrics---being able to graph the average latency using 1 second
intervals, but also rolling up the sums and counts to graph the average latency over
much larger time periods using much larger intervals perfectly accurately.


Unfortunately, the latencies of web requests are usually extremely skewed---the
median response time might be in the milliseconds whereas there could be a
couple of outlying responses that take minutes (Figure~\ref{fig:histograms}).  A
simple average, while easy to monitor can be easily skewed by outlying values as
can be seen in Figure~\ref{fig:p50meanp75}.

As the average response time is not a particularly useful measure, we are
instead interested in tracking quantiles such as the 50th and the 99th
percentiles (we will also refer to these as the \emph{p50} and \emph{p99}).
The ability to compute quantiles over aggregated metrics has been recognized to
be an essential feature of any monitoring system~\cite{DeanBarroso-2013}.

Quantiles are famously impossible to compute exactly without holding on to all
the data~\cite{MunroPaterson-1980}. If one wanted to track the median request
latency over time for a web application that is handling millions of requests a
second, this would mean sending millions of data points to the monitoring
service which could then calculate the median by sorting the data. If one wanted
the median aggregated over longer time intervals the monitoring service would
have to store all these data points and then calculate the median over the larger
set of points.

Given how expensive calculating exact quantiles can be for both storage and
network bandwidth, most monitoring systems will compress the data into sketches
and compute approximate quantiles. More formally, given a multiset $S$ of size
$n$ over $\RR$, the \emph{$q$-quantile} item $x_q \in S$ is the item $x$ whose
rank $R(x)$ in the sorted multiset $S$ is $\floor{1+q(n-1)}$ for $0 \leq q \leq
1$, where the rank $R(x)$ is the number of elements in $S$ smaller than or equal
to $x$.\footnote{This definition of quantile is also known as the \emph{lower
    quantile}. Replacing the floor with a ceiling gives us what's known as the
  \emph{upper quantile}.}  Some special quantiles include $x_1$, the maximum
element of $S$, and the median $x_{0.5}$.  There has been a long line of work on
sketching data streams so that the \emph{rank accuracy} is preserved, \ie for
any value $v$, the sketch provides an estimate rank $\tilde{R}$ such that
$\abs{\tilde{R}(v) - R(v)} \leq \eps n$ (see \cite{LWYC-2016} and
\cite{GreenwaldKhanna-2016} for excellent surveys on much of this work as well
as additional motivation for sketching quantiles).

Unfortunately, for data sets with heavy tails, rank-error guarantees can return
values with large relative errors. Consider again the histogram
of 2M request response times in Figure~\ref{fig:histograms}.
If we have a quantile sketch with a rank accuracy of 0.005, and ask for the 99th
percentile, we are guaranteed to get a value between the 98.5th and 99.5th
percentile. In this case this is anywhere from 2 to 20 seconds, which from an
end-user's perspective is the difference between an annoying delay and giving up
on the request.

Given the inadequacy of rank accuracy for tracking the higher order quantiles
for distributions with heavy tails, we turn instead to relative accuracy.
\begin{definition}
  $\tilde{x}_q$ is an \emph{$\alpha$-accurate $q$-quantile} if
  $\abs{\tilde{x}_q - x_q} \leq \alpha x_q$ for a given $q$-quantile item
  $x_q\in S$. We say that a data structure is an \emph{$\alpha$-accurate
    $(q_0,q_1)$-sketch} if it can output $\alpha$-accurate $q$-quantiles for all
  $q_0 \leq q \leq q_1$.
\end{definition}

\begin{figure*}
  \centering
  \includegraphics[width=.8\textwidth]{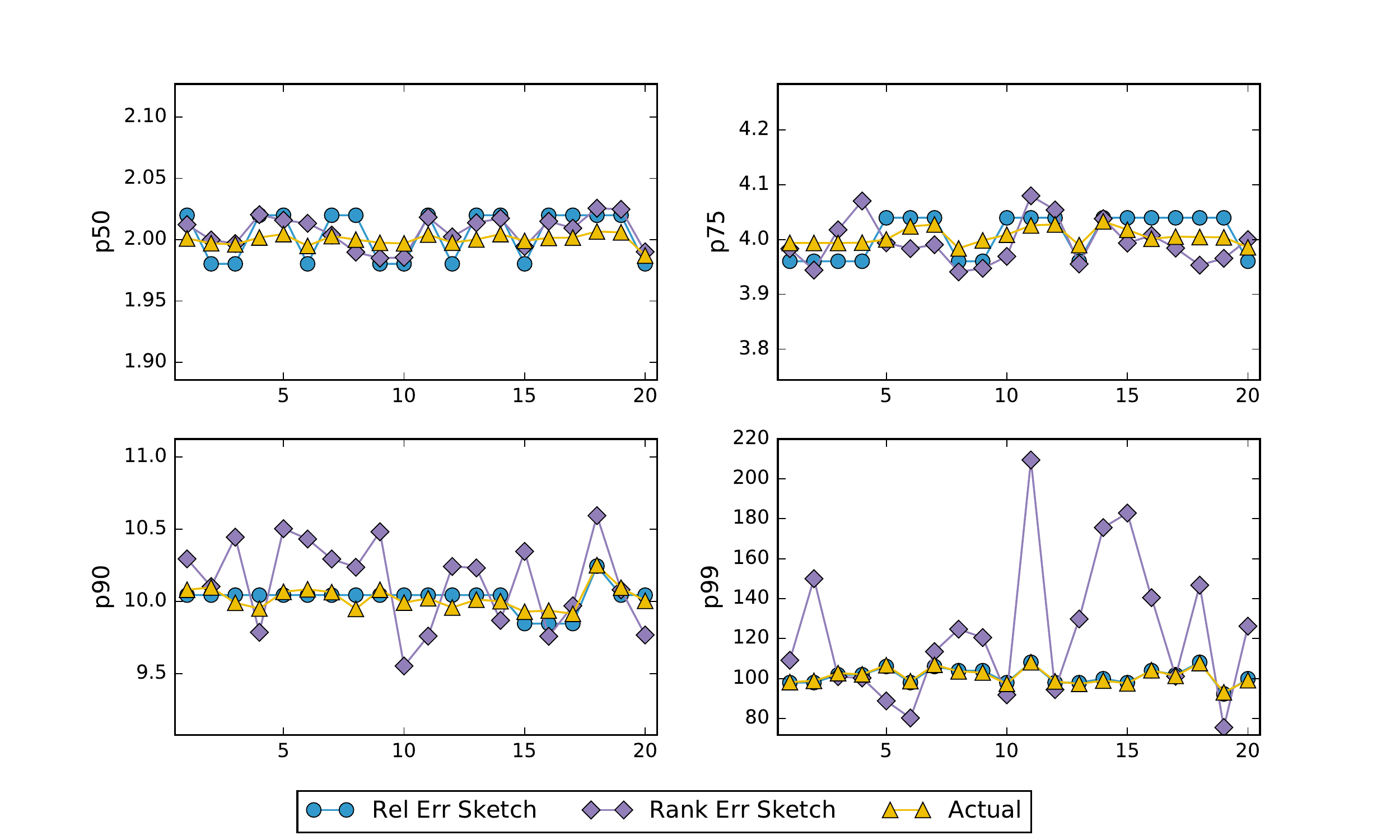}
  \caption{Actual quantiles vs. $0.005$-rank-accurate and
    $0.01$-relative-accurate quantiles of a data stream (20 batches of 100,000
    values).}
  \label{fig:timeseries}
\end{figure*}

To further illustrate the difference between rank accuracy and relative accuracy
consider Figure~\ref{fig:timeseries}. The graphs show the actual p50, p75, p90
and p99 values along with the quantile estimates from a sketch with $0.005$
rank accuracy and a sketch with $0.01$ relative accuracy.

\subsection{Our Results}
In Section~\ref{sec:sketch} we describe our relative-error sketch, dubbed the
Distributed Distribution Sketch (DDSketch), and we discuss different
implementation strategies. In Section~\ref{sec:bounds} we prove that the sketch
can handle data that is as heavy-tailed as that which comes from a distribution
whose logarithm is subexponential with parameters $(\sigma, b)$, which includes
heavy-tailed distributions such as the log-normal and Pareto distributions. We
show that for the Pareto distribution, the size of an $\alpha$-accurate
$(o(1),1)$-sketch is: \[O\left(\frac{b \log n/\delta}
{\log(\paren{1+\alpha}/\paren{1-\alpha})}\right)\] with probability
$1-\delta$. (Note that our results hold for data coming from any distribution
without any independence assumptions as long as the tail of the empirical data
is no larger than that for a Pareto distribution).  In Section~\ref{sec:exp} we
present our experimental results.

\subsection{Related Work}\label{subsec:related}

Quantile sketching dates back to 1980 when Munro and Paterson
\cite{MunroPaterson-1980} demonstrated the first quantile sketching algorithm
with formal guarantees. The best known rank-error quantile sketch is that of
Greenwald and Khanna \cite{GreenwaldKhanna-2001} whose deterministic sketch (GK)
provides $\eps$ rank accuracy using $O((1/\eps)\log(n\eps))$ space.

In addition to accuracy and size, a desirable property of a sketch is
\emph{mergeability} \cite{ACHPWY-2012}. That is, several sketches of different
data sets can be combined into a single sketch that can accurately answer
quantile queries over the entire data set. Mergeability has increasingly become
a necessary property as systems become more distributed. Equi-depth histograms
\cite{CGHJ-2011} are a good example of non-mergeable data set synopses as there
is no way to accurately combine overlapping buckets. GK is only known to be
``one-way'' mergeable, that is the merging operation itself can not be
distributed.

There is a line of work using randomness culminating in a rank-error quantile
sketch that uses only $O((1/\eps)\log\log(1/\delta))$ space (where $\delta$ is
the probability of failure) \cite{KLL-2016} with full mergeability. However, all
of the above solutions, deterministic or randomized, have high relative error
for the larger quantiles on heavy-tailed data (in practice we have found it to
be worse for the randomized algorithms).

The problems of having high relative errors on the larger quantiles has been
addressed by a line of work that still uses rank error, but promises lower rank
error on the quantiles further away from the median by biasing the data it keeps
towards the higher (and lower) quantiles \cite{CKMS-2005}, \cite{CKMS-2006},
\cite{DunningErtl-2019}. The latter, dubbed $t$-digest, is notable as it is one
of the quantile sketch implementations used by Elasticsearch
\cite{Elasticsearch}. These sketches have much better accuracy (in rank) than
uniform-rank-error sketches on percentiles like the p99.9, but they still have
high relative error on heavy-tailed data sets. Like GK they are only one-way
mergeable.

The only relative-error sketch in the literature to our knowledge is the HDR
Histogram \cite{Tene} (and is the other quantile sketch implementation used by
Elasticsearch). It has extremely fast insertion times (only requiring low-level
binary operations), as the bucket sizes are optimized for insertion speed
instead of size, and it is fully mergeable (though very slow). The main downside
for HDR Histogram is that it can only handle a bounded (though very large) range
that might not be suitable for certain data sets. It also has no published
guarantees, though much of the analysis we present for DDSketch can be made to
apply to a version of HDR Histogram that more closely resembles DDSketch with
slightly worse guarantees.

A recent quantile sketch, called the Moments sketch~\cite{GDTSB-2018} takes an
entirely different approach by estimating the moments of the underlying
distribution. It has notably fast merging times and is fully mergeable. The
guaranteed accuracy, however, is only for the \emph{average} rank error
$\tilde{\eps}$, unlike all the sketches above which have guarantees for the
worst-case error (whether rank or relative). The associated size bound is
$O(1/\tilde{\eps})$. In practice, the sketch also has a bounded range as the
moments quickly grow larger, and they will eventually cause floating point
overflow errors.

We compare the performance of DDSketch to GK, HDR, and Moments in
Section~\ref{sec:exp}.

\begin{table}[ht]
\centering
\caption{Quantile Sketching Algorithms}
\begin{tabular}{c c c c }
  \hline
  & guarantee & range & mergeability\\
  \hline
DDSketch & relative & arbitrary & full \\
\hline
HDR Histogram & relative & bounded & full \\
\hline
GKArray & rank & arbitrary & one-way\\
\hline
Moments & avg rank & bounded & full\\
\hline
\end{tabular}
\label{table:sketch_algos}
\end{table}

A related line of work exists in constructing histograms (see \cite{CGHJ-2011}
for a thorough survey). The accuracy of a histogram is measured using the
distance between the actual values and the values of the buckets to which the
original values are assigned. The task is to find the histogram with $B$ buckets
that minimizes the overall distance. Optimal algorithms \cite{JKMPSS-1998} use
dynamic programming and are usually considered to be too costly, and thus
approximation algorithms are often considered. The most popular distance in the
literature is the squared L2 distance (such a histogram is called the
\emph{v-optimal} histogram), but relative-error approximation algorithms exist
as well~\cite{GSW-2004}, \cite{GKS-2006} (though these algorithms use $\Omega(n)$
space).

Note that while one can try to use these histograms to answer quantile queries,
there are no guarantees on the error of any particular quantile query, as the
only error guarantees are global and not for any individual item. Moreover, the
error guarantees are always relative to an unknown optimal (for the number of
buckets) solution, not an absolute error guarantee. There is also no
straightforward way to merge histograms as the bucket boundaries are based on
the data, which can be wildly different for each histogram.

\section{DDSketch}\label{sec:sketch}

We will first describe the most basic version of our algorithm that will be able
to give $\alpha$-accurate $q$-quantiles for any $0 \leq q \leq 1$. It is
straightforward to insert items into this sketch as well as delete items and
merge sketches. Then we will show how to modify the sketch so that it gives
$\alpha$-accurate $q$-quantiles for $q_0 \leq q \leq 1$ with bounded size.
Section~\ref{subsec:implementation} will go over various implementation
options for the sketch.

\subsection{Sketch Details}
Let $\gamma := \paren{1+\alpha}/\paren{1-\alpha}$.
The sketch works by dividing $\RR_{>0}$ into fixed
buckets. We index our buckets by $i \in \ZZ$, and each bucket $B_i$ counts the
number of values $x$ that fall between: $\gamma^{i-1} < x \leq \gamma^i$. That
is, given a value $x$ we will assign it to the bucket indexed by
$\ceil{\log_\gamma(x)}$:

\begin{algorithm}[ht]
  \caption{Insert($x$)}
  \BlankLine
  \KwIn{$x \in \RR_{>0}$}
  \BlankLine
  $i \leftarrow \ceil{\log_\gamma(x)}$\;
  $B_i \leftarrow B_i + 1$\;
\end{algorithm}

Deletion works similarly. Since the bucket boundaries are independent of the
data, any two sketches using the same value for $\gamma$ can be merged by simply
summing up the buckets that share an index.

A simple lemma shows that every value gets assigned to a bucket whose boundary
values are enough to return a relative-error approximation to its value.

\begin{lemma}\label{lem:easy}
  For a given $q$-quantile item $x_q\in S$ and bucket index $i=\ceil{\log_\gamma(x_q)}$,
  let $\tilde{x}_q = 2\gamma^i/\paren{\gamma+1}$.
  Then $\tilde{x}_q$ is an $\alpha$-accurate $q$-quantile.
\end{lemma}
\begin{proof}
  Note that by definition of $\gamma$:
  \[\alpha = 1-\frac{2}{\gamma+1} = \frac{2\gamma}{\gamma+1}-1\]
  Moreover, $\gamma^{i-1} < x_q \leq \gamma^{i}$. So if $x_q \geq \tilde{x}_q$, then:
  \[\frac{x_q-\tilde{x}_q}{x_q} = 1-\frac{\tilde{x}_q}{x_q} \leq
  1-\frac{\tilde{x}_q}{\gamma^i} = 1-\frac{2\gamma^i}{\gamma^i\paren{\gamma+1}}
  = \alpha\]
  Similarly if $x_q \leq \tilde{x}_q$, then:
  \[\frac{\tilde{x}_q-x_q}{x_q} = \frac{\tilde{x}_q}{x_q}-1 <
  \frac{\tilde{x}_q}{\gamma^{i-1}}-1 =
  \frac{2\gamma^i}{\gamma^{i-1}\paren{\gamma+1}}-1 = \alpha\]
  Combining both cases:
  \[\Abs{x_q-\tilde{x}_q} \leq \alpha x_q. \]
\end{proof}

To answer quantile queries, the sketch sums up the buckets until it finds the bucket containing the $q$-quantile value $x_q$:

\begin{algorithm}[ht]
  \caption{Quantile($q$)}
  \BlankLine
  \KwIn{$0 \leq q \leq 1$}
  $i_0 \leftarrow \min(\set{j:B_j > 0})$;
  \BlankLine
  $count \leftarrow B_{i_0}$\;
  $i \leftarrow i_0 $\;
  \While{$count \leq q(n-1)$}{
    $i \leftarrow \min(\set{j:B_j > 0 \land j > i})$\;
    $count \leftarrow count + B_i$;
  }
  \Return{$2\gamma^i/\paren{\gamma+1}$}\;
\end{algorithm}

Given Lemma~\ref{lem:easy}, the following Proposition easily follows:
\begin{proposition}
Given $\alpha > 0$ and $0 \leq q \leq 1$, Quantile($q$) return an $\alpha$-accurate $q$-quantile.
\end{proposition}
\begin{proof}
  Let's refer to the ordered elements of the multiset $S$ as $x_{(1)} \leq \dots \leq x_{(n)}$, so that by definition of the quantile, $x_q = x_{(\floor{1+q(n-1)})}$. We will also write $c(x)$ the number of elements in $S$ that are less than or equal to $x$. Note that for any $j$, we always have $c(x_{(j)}) \geq j$. Quantile($q$) outputs $2\gamma^i/\paren{\gamma+1}$ where $i = \min(\set{j:c(\gamma^{j}) > q(n-1)})$. Given Lemma~\ref{lem:easy}, it is enough to prove that $i = \ceil{\log_\gamma{x_{(\floor{1+q(n-1)})}}}$.

  Let $k$ be the largest integer so that $x_{(k)} \leq \gamma^i$. It is clear that $x_{(k)}$ is in the bucket of index $i$, so that $i = \ceil{\log_\gamma{x_{(k)}}}$. By definition of $k$, $k = c(x_{(k)})$ and, because there is no element of $S$ in the bucket of index $i$ that is greater than $x_{(k)}$, we also know that $c(x_{(k)}) = c(\gamma^i)$. Thus, $k = c(x_{(k)}) = c(\gamma^i) > q(n-1)$ and, given that $k$ is an integer, $k \geq \floor{1+q(n-1)}$ follows.
  Therefore,
  $i = \ceil{\log_\gamma x_{(k)}} \geq \ceil{\log_\gamma x_{(\floor{1+q(n-1)})}}$.

  By contradiction, if $i > \ceil{\log_\gamma x_{(\floor{1+q(n-1)})}}$, then
  $i-1 \geq \ceil{\log_\gamma x_{(\floor{1+q(n-1)})}}$ and $\gamma^{i-1} \geq
  x_{(\floor{1+q(n-1)})}$. As a consequence:
   \[c(\gamma^{i-1}) \geq c(x_{(\floor{1+q(n-1)})}) \geq \floor{1+q(n-1)} > q(n-1),\]
   which violates the definition of $i$. Hence,
   \[i = \ceil{\log_\gamma{x_{(\floor{1+q(n-1)})}}},\]
   and the result follows.
\end{proof}

However buckets are stored in memory (e.g., as a dictionary that maps indices to
bucket counters, or as a list of bucket counters for contiguous indices), the
memory size of the sketch is at least linear in the number of non-empty buckets.
Therefore, a down-side to the basic version of DDSketch is that for worst-case
input, its size can grow as large as $n$, the number of elements inserted into it. A
simple modification will allow us to guarantee logarithmic size bounds for
non-degenerate input, and Section~\ref{sec:bounds} will show that the
modification will never affect the ability to answer $q$-quantile queries
for any constant $q$.

The full version of DDSketch is a simple modification that addresses its
unbounded growth by imposing a limit of $m=f(n)$ on the number of buckets it
keeps track of. It does so by collapsing the buckets for the smallest indices:

\begin{algorithm}[ht]
  \caption{DDSketch-Insert($x$)}
  \BlankLine
  \KwIn{$x \in \RR_{>0}$}
  \BlankLine
  $i \leftarrow \ceil{\log_\gamma(x)}$\;
  $B_i \leftarrow B_i + 1$\;
  \If{$\Abs{\set{j:B_j > 0}} > m$}{
    $i_0 \leftarrow \min(\set{j:B_j > 0})$\;
    $i_1 \leftarrow \min(\set{j:B_j > 0 \land j > i_0})$\;
    $B_{i_1} \leftarrow B_{i_1} + B_{i_0}$\;
    $B_{i_0} \leftarrow 0$\;
    }
\end{algorithm}

Given that our sketch has predefined bucket boundaries for a given $\gamma$,
merging two sketches is straightforward. We just increase the counts of the
buckets for one sketch by those of the other. This, however, might increase the
size of the sketch beyond the limit of $m=f(n)$, where $n$ is now the number of
elements in the resulting merged sketch. As with the insertion, we stay within
the limit by collapsing the buckets with smallest indices:

\begin{algorithm}[ht]
  \caption{DDSketch-Merge($S'$)}
  \BlankLine
  \KwIn{DDSketch $S'$}
  \BlankLine
  \ForEach{$i:B_i > 0 \lor B'_i > 0$}{
    $B_i \leftarrow B_i + B'_i$\;
  }
  \While{$\Abs{\set{j:B_j > 0}} > m$}{
    $i_0 \leftarrow \min(\set{j:B_j > 0})$\;
    $i_1 \leftarrow \min(\set{j:B_j > 0 \land j > i_0})$\;
    $B_{i_1} \leftarrow B_{i_1} + B_{i_0}$\;
    $B_{i_0} \leftarrow 0$\;
  }
\end{algorithm}

We trade off the benefit of a bounded size with not being able to correctly
answer $q$-quantile queries if $x_q$ belongs to a collapsed bucket. The next
lemma shows a sufficient condition for a quantile $q$ to be $\alpha$-accurately
answered by our algorithm:

\begin{proposition}\label{prop:acc}
  DDSketch can $\alpha$-accurately answer a given $q$-quantile query if:
\[x_1 \leq x_q \gamma^{m-1} .\]
\end{proposition}
\begin{proof}
For any particular quantile $q$, $x_q$ will be $\alpha$-accurate as long as it
belongs to one of the $m$ buckets kept by the sketch. Let's refer to that bucket
index as $i_q$, which holds values between $\gamma^{i_q-1}$ and
$\gamma^{i_q}$. If the maximum bucket (that holds $x_1$) has index
$i_1 \leq i_q + m - 1$, then the bucket $i_q$ has definitely not been collapsed.
Thus, given that $x_1 \leq x_q \gamma^{m-1}$, then
$\gamma^{i_1-1}< x_1 \leq x_q\gamma^{m-1} \leq \gamma^{i_q}\gamma^{m-1}$,
and $i_1-1 < i_q +m-1 $, which is equivalent to
$i_1\leq i_q +m -1$ as these indices are integers.
\end{proof}

We'll discuss the trade-offs between the accuracy $\alpha$, the minimum accurate
quantile $q$, the number of items $n$, and the size of the sketch $m$ in
Section~\ref{sec:bounds}.

\subsection{Implementation Details}\label{subsec:implementation}
Most systems often have built-in timeouts and a minimum granularity, so the
values coming into a sketch usually have an effective minimum and maximum.
Importantly, our sketch does not need to know what those values are beforehand.

It is straightforward to extend DDSketch to handle all of $\RR$ by keeping a
second sketch for negative numbers. The indices for the negative sketch need to
be calculated on the absolute values, and collapses start from the highest
indices.

Like most sketch implementations, it is useful to keep separate track of the
minimum and maximum values. Given how buckets are defined for DDSketch, we also
keep a special bucket for zero (and all values within floating-point error of
zero when calculating the index, which involves computing the logarithm of the
input value).

The size of the sketch can be set to grow by setting $m = c \log n$, which will
match the upper bounds discussed in Section~\ref{sec:bounds}, but in practice
$m$ is usually set to be a constant large enough to handle a wide range of values.
As an example, for $\alpha=0.01$, a sketch of size 2048 can handle values from
80 microseconds to 1 year, and cover \emph{all} quantiles.

If $m$ is set to a constant, it often makes sense to preallocate the sketch
buckets in memory and keep all the buckets between the minimum and maximum
buckets (perhaps implemented as a ring buffer). If the sketch is allowed to grow
with $n$, then the sketch can either grow with every order of magnitude of $n$,
or one can implement the sketch in a sparse manner so that only buckets with
values greater than 0 are kept (sacrificing speed for space efficiency).

\section{Distribution Bounds}\label{sec:bounds}

For most practical applications, \eg tracking the latency of web requests to a
particular endpoint, one cares about constant quantiles such as those
around the median such as 0.25, 0.5, 0.75, or those towards the edge such as
0.9, 0.95, 0.99, 0.999. Thus by Proposition~\ref{prop:acc}, we will focus on the
necessary conditions for $x_1 \leq x_{q} \gamma^{m-1}$ or:
\begin{equation}\label{eqn:maxmdn}
  \frac{\log(x_1)-\log(x_{q})}{\log(\gamma)} +1 \leq m.
\end{equation}
for $q=\Theta(1)$, though our results will apply for $q = \Omega(1/\sqrt{n})$.
(For simplicity, we will assume that $qn$ is a whole number in this section.)

For any fixed $\gamma$ and $m$, it is easy to come up with a data set $S$ for
which the condition does not hold, \eg $S = \set{\gamma^1, \gamma^2, \ldots,
  \gamma^{2m} }$.  Given that distribution-free results are impossible, for our
formal results, we will instead assume that our data is i.i.d.\ from a
particular family of distributions, and then show how large the sketch would
have to be for Equation~\ref{eqn:maxmdn} to hold. We are able to obtain good
bounds for families of distributions as general as those whose logarithms are
subexponential (e.g., Pareto distributions). While our bounds are obtained by
assuming i.i.d., in practice, as long as the tail of the empirical distribution
is no fatter than that of a Pareto, we do not need to assume anything about the
data generating process at all.

We will bound the LHS of Equation~\ref{eqn:maxmdn} by showing that with
probability greater than $1-\delta_1$ the sample quantile $x_{q}$ is greater
than a quantile just below it. Then we will show that with probability greater
than $1-\delta_2$, the sample maximum is less than some bound. Finally by the
union bound we get that the probability that both bounds apply is greater than
$1-\delta_1-\delta_2$. In practice, the probability of failing these bounds is
smaller as the sample maximum and sample quantiles become quickly independent as
$n$ grows.

\subsection{Sample Quantiles}

Let capital $X_1,\ldots,X_n$ denote $n$ independent real-valued random
variables drawn from a distribution with cdf
$F:\RR\rightarrow [0,1]$. The generalized inverse of $F$, $F\inv(p)$
is known as the quantile function of $F$. Let $X_{(1)} \leq
X_{(2)}\leq\cdots\leq X_{(n)}$ denote the \emph{order statistics} of
$X_1,\ldots, X_n$ (\ie the ordered random variables).

The next Lemma shows that with high probability a lower sample quantile can't
fall too far below the actual quantile.
\begin{lemma}\label{lem:percent}
  Let $X_{(1)} \leq X_{(2)}\leq\cdots\leq X_{(n)}$ be the order statistics for
  i.i.d. random variables $X_i$ distributed according to $F$. Let
  $t = \sqrt{\log(1/\delta_1)/2n}$ and $t < q \leq 1/2$, then
  \[\Pr\Brac{X_{(q n)} > F\inv(q - t)} \geq 1-\delta_1.\]
\end{lemma}
\begin{proof}
  The proof follows Chv\'atal's proof of a special case of the Hoeffding bound
  \cite{Chvatal-1979}.  For any single random variable $X$ drawn from a
  distribution with cdf $F$, $\Pr\brac{X \leq F\inv(p)} = p$. Then for any
  particular order statistic and any $x \geq 1$:
  \begin{align}
    & \Pr\Brac{X_{(k)} \leq F\inv(p)}\nonumber\\
    & = \sum_{i=k}^{n}\binom{n}{i}p^i(1-p)^{n-i}  \nonumber\\
    & \leq \sum_{k}^{n}\binom{n}{i}p^i (1-p)^{n-i} x^{i-k} +
    \sum_{0}^{k-1}\binom{n}{i}p^i (1-p)^{n-i} x^{i-k}  \nonumber\\
    & = x^{-k} \sum_{0}^{n}\binom{n}{i}p^i(1-p)^{n-i} x^{i} = x^{-k} (px+(1-p))^n  \label{eqn:binomial}
  \end{align}
  where the last equality is by the Binomial Theorem.

  Equation~\ref{eqn:binomial} is minimized when $x = (1-p)k/p(n-k)$, and taking
  $k = q n$, our bound becomes:
  \begin{equation}\label{eqn:minimized}
    \Paren{\frac{p(n-k)}{(1-p)k}}^k\Paren{\frac{(1-p)n}{n-k}}^n =
    \Paren{\frac{p }{ q }}^{q n}\Paren{\frac{(1-p)}{(1-q)}}^{n(1-q)}
  \end{equation}
  Note that for $x\geq 1$, $p\leq q \leq 1/2$, and that the
  bound is trivial when $p=q$. However, if we take $p=q-t$, we get:
  \begin{align*}
    (\ref{eqn:minimized})
    & = \exp \Paren{-q n \log\Paren{\frac{q}{q-t}} - \Paren{1-q}n \log\Paren{\frac{1-q}{1-q+t}}}\\
    & =  \exp \Paren{-n \int_{q-t}^q\Paren{\frac{q}{x}  -\frac{1-q}{1-x} }dx}\\
    & =  \exp \Paren{-n \int_{q-t}^q\frac{q-x}{x(1-x)} dx}\\
    & \leq \exp\Paren{-4n \int_{q-t}^q(q-x) dx} =  \exp\Paren{-2n t^2} = \delta_1
  \end{align*}
  for $t = \sqrt{\log(1/\delta_1)/2n}$, and where the last inequality uses the
  fact that $x(1-x) \leq 1/4$ for all $x\in \RR$.
\end{proof}

\begin{figure*}[ht]
  \centering
  \includegraphics[width=\textwidth]{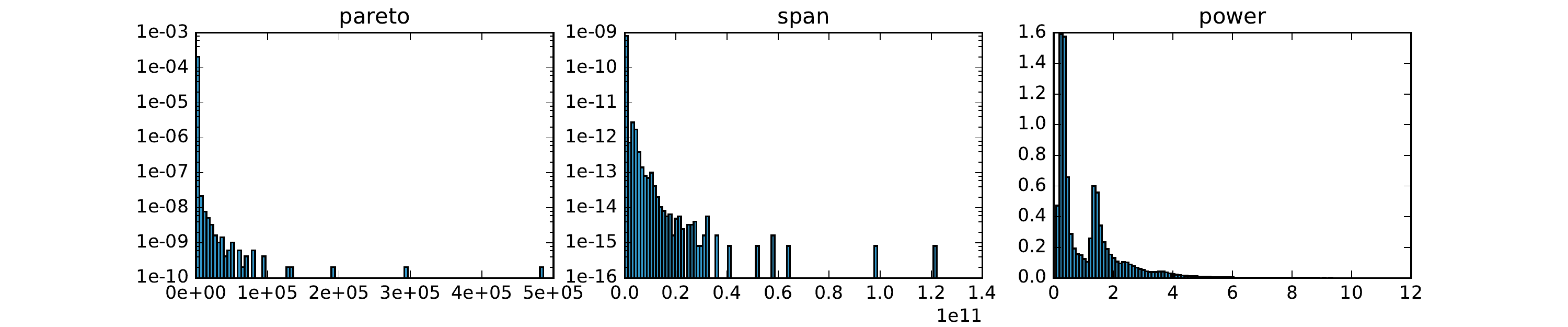}
  \caption{Histograms of the \emph{pareto}, \emph{span} and \emph{mpcat-obs} data sets.
  The $y$-axes of the \emph{pareto} and \emph{span} data sets are plotted on log scales due to their heavy-tailed nature.}
  \label{fig:datadist}
\end{figure*}

\subsection{Sample Maximums}

We will first bound the sample maximum for subexponential
distributions, which include the Gaussian, logistic, chi-squared,
exponential, and many others.

\begin{definition}
  A random variable $X\in \RR$ is said to be \emph{subexponential} with parameters $(\sigma, b)$ if
  \[\EE\Brac{\exp\Paren{\lambda(X-\EE X)}} \leq \exp\Paren{\sigma^2 \lambda^2/2},\]
  for $0\leq \lambda \leq 1/b$.
\end{definition}

Using Chernoff-type techniques, one can obtain concentration
inequalities for subexponential variables~\cite{BuldyginKozachenko-2000}.
\begin{theorem}\label{thm:concentration}
  Let $X$ be a subexponential random variable with parameters $(\sigma, b)$. Then,
  \[\Pr\Brac{ X - \EE X \geq t} \leq \exp\Paren{-t^2/2\sigma^2}\]
  for $0\leq t \leq \sigma^2/ b$, and
  \[\Pr\Brac{ X - \EE X \geq t} \leq \exp\Paren{-t/2b}\]
  for $t > \sigma^2/ b$.
\end{theorem}

Now we can lower-bound the sample maximum by the complement of the event that
none of the sample is greater than $t$.
\begin{corollary}\label{cor:maximum}
  Let $X_{(1)} \leq X_{(2)}\leq\cdots\leq X_{(n)}$ be the order statistics for
  i.i.d. subexponential random variables $X_i$ with parameters $(\sigma, b)$,
  and $t = 2b \log(n/ \delta_2)$. Then the sample maximum is less than $t$ with
  probability at least $1-\delta_2$.
\end{corollary}
\begin{proof}
  By Theorem~\ref{thm:concentration},
  \begin{align*}
    \Pr\Brac{X_{(n)} - \EE X > t} &< 1 - \Paren{1- e^{-t/2b}}^n\\
    &= 1 - \Paren{1-\frac{\delta_2}{n}}^n  <  \delta_2,
  \end{align*}
  where the final inequality is by Bernoulli's inequality.
\end{proof}

\subsection{Sketch Size Bounds}\label{subsec:sizebounds}
For subexponential distributions, we can bound Equation~\ref{eqn:maxmdn} by
combining Lemma~\ref{lem:percent} and Corollary~\ref{cor:maximum}:

\begin{theorem}\label{thm:sizebound}
  Let $X_{(1)} \leq X_{(2)}\leq\cdots\leq X_{(n)}$ be the order statistics for
  i.i.d. random variables $X_i$ distributed according to a subexponential
  distribution $F$ with parameters $(\sigma, b)$.
  Then with probability at least $1-\delta_1-\delta_2$, DDSketch is an
  $\alpha$-accurate $(q,1)$-sketch with size at most
  $\paren{\log X_{(n)}-\log X_{(qn)}}/\log(\gamma)+1$, which is bounded from above by:
  \[
  \frac{\log\Paren{2b \log\Paren{n/\delta_2} + \EE X} -\log\Paren{F\inv\Paren{q-t}} }{\log(\gamma)} + 1
  \]
  for $\gamma = \paren{1+\alpha}/\paren{1-\alpha}$, $t = \sqrt{\log(1/\delta_1)/2n}$, and $t < q \leq 1/2$.
\end{theorem}

\medskip

\noindent\textbf{Exponential.} For concreteness, let's take $\delta_1 = \delta_2
= e^{-10}$ and $\alpha = 0.01$ (i.e., $\gamma \approx 1.02$), and let's consider the
exponential distribution with cdf $F(t; \lambda) = 1- \exp(-\lambda t)$ for $t
\geq 0$, and 0 otherwise. The exponential distribution is subexponential with
parameters $(2/\lambda, 2/\lambda)$.

If $n > 320$, then $p > 3/8$, and the sample median is at least $F\inv\Paren{p}
> - \lambda\inv\log(1-3/8) > 0.47/\lambda$. The sample maximum\footnote{The
  factor of 4 can be removed from the bound for the sample maximum if we analyze
  the exponential distribution directly instead of using the generic bounds for
  subexponential distributions.} is at most $4\lambda\inv(\log(n) + 10 + 1/4)$,
and $1/\log(\gamma) < 51$ so we can bound the size from
Theorem~\ref{thm:sizebound} by: $51(\log(4\log n+41)-\log(0.47)) +1$.

This means that even with a sketch of size 273 one can 0.01-accurately maintain
the upper half order statistics of over a million samples with probability
greater than 0.99991. This grows double-exponentially, so a sketch of size 1000
can 0.01-accurately maintain the upper half order statistics of over
$\exp(\exp(17))$ values with that same probability.

\medskip

\noindent\textbf{Pareto.} The double logarithm in our size bound from
Theorem~\ref{thm:sizebound} allows us to handle distributions with much fatter
tails as well. The Pareto distribution distribution has cdf $F(t; a, b) = 1-
(b/t)^a$. If $X$ is a random variable drawn from this distribution, then
$Y=\log(X/b)\sim Exp(a)$. Thus, we can reuse the arguments above to get that
with probability at least $1-\delta_1-\delta_2$:
\[\log(X_{(n)}) < 4a\inv\log(n  / \delta_2) + a\inv + \log(b)\]
and
\[\log(X_{(n/2)}) > -a\inv\log(1/2+\sqrt{\log(1/\delta_1)/2n}) +\log(b).\]

As before, let's take $\delta_1 = \delta_2 = e^{-10}$, $\alpha = 0.01$, and
assume that $n>320$. With probability greater than $1-e^{-10}-e^{-10} \approx
0.99991$:

\[  \frac{\log(X_{(n)})-\log(X_{(n/2)})}{\log(\gamma)}+1  < 51 a \inv(4\log n+11) +1\]

Given that Pareto distributions have exponentially fatter tails than exponential
distributions, the sketch size upper bounds increase accordingly. Taking $a = 1$,
this means that we require a sketch of size 3380 to 0.01-accurately maintain the
upper half order statistics of over a million samples with probability greater
than 0.99991. A sketch of size 10000, can 0.01-accurately maintain the upper
half order statistics of over $\exp(46)$ values with that same probability.

\medskip

\noindent\textbf{Other Distributions.} We focused on subexponential tails and
the Pareto distribution in this section as we believe it to best represent the
worst case for practical use-cases of quantile sketching. For lighter tails
such as subgaussians and thus for lognormal distributions, we can of course get
much tighter bounds.

\begin{figure*}[!ht]
\centering
\includegraphics[width=\textwidth]{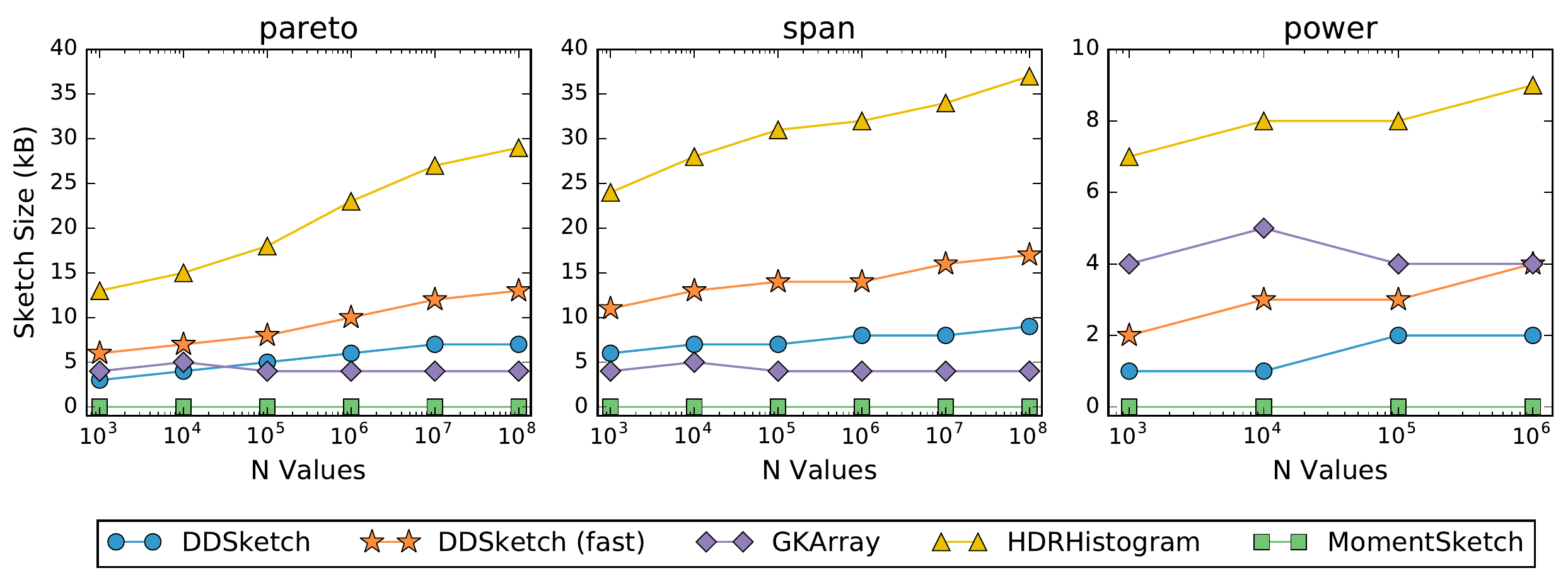}
\caption{Sketch size in memory in kB.}
\label{fig:sketchsize}
\end{figure*}

\section{Evaluation}\label{sec:exp}

We provide implementations of DDSketch in Java~\cite{Implementation-java},
Go~\cite{Implementation-go} and Python~\cite{Implementation-py}. Our Java
implementation provides multiple versions of DDSketch: buckets can be stored in
a contiguous way (for fast addition) or in a sparse way (for smaller memory
footprint). The number of buckets can grow indefinitely or be bounded with a
fixed maximum of $m$ buckets, collapsing the buckets of lowest or highest
indices. The mapping of the values to their bucket indices can be logarithmic,
as defined above, but we also provide alternative mappings that are faster to
evaluate while still ensuring relative accuracy guarantees. Those mappings make
the most of the binary representation of floating-point values, which provides a
costless way to evaluate the logarithm to the base 2. In between a linear or
quadratic interpolation can be used so that the logarithm to any base can be
approximated. Those mappings define buckets whose sizes are not optimal under
the constraint of ensuring relative accuracy guarantee as some of them are
smaller than necessary. Their faster evaluation than the memory-optimal
logarithmic mapping comes at the cost of requiring more buckets to cover a given
range of values, and therefore a memory footprint overhead in DDSketch. We refer
to this version of the code as DDSketch (fast) in our experiments.

We compare DDSketch against the Java implementation \cite{Tene} of HDR
Histogram, our Java implementation of the GKArray version of the GK sketch
\cite{Implementation-java}, as well as the Java implementation of the Moments
sketch \cite{Moments-sketch-impl} (all three discussed in Section~\ref{subsec:related}). HDR Histogram is a
relative-error sketch for non-negative values. Its accuracy is expressed as the
number of significant decimal digits $d$ of the values. GKArray guarantees that
the rank error of the estimated quantiles will be smaller than $\epsilon$ after
adding values. The Moments sketch has guarantees on the average rank error
bounded by the number of moments $k$ that are estimated.

We consider three data sets, and compare the size in memory of the sketches, the
speed of adding values and merging sketches, and the accuracy of the estimated
quantiles. The measurements are performed with the Java implementations of all
four sketches.

The sketch parameters are chosen so that the targeted relative accuracy for
DDSketch and HDR Histogram is 1\%. For GKArray, we use a rank accuracy that
gives roughly similar memory footprints as DDSketch. For the Moments sketch, we
use the maximum recommended numbers of moments, as per the Java implementation
documentation, and we also use the arcsinh transform (called \emph{compression} in the
code), which makes the sketch more accurate for distributions with heavy tails.
Those parameters are summarized in Table~\ref{table:params}.

\begin{figure}[!ht]
  \centering
  \includegraphics[width=\columnwidth]{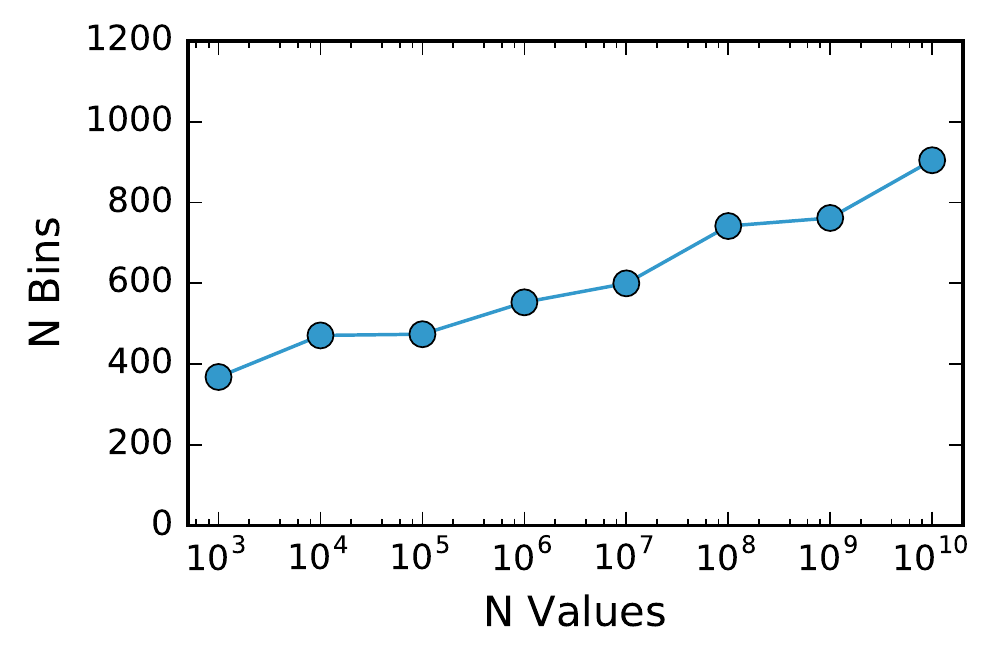}
  \caption{Number of bins in DDSketch for the \emph{pareto} data set.}
  \label{fig:nbins}
\end{figure}

\begin{table}[ht]
\centering
\caption{Experiment Parameters}
\begin{tabular}{c c}
\hline
\multirow{2}{*}{DDSketch}
& $\alpha = 0.01$ \\
& $m = 2048$ \\
\hline
HDR Histogram & $d = 2$ \\
\hline
GKArray & $\epsilon$ = 0.01 \\
\hline
\multirow{2}{*}{Moments sketch}
& $k$ = 20 \\
& compression enabled \\
\hline
\end{tabular}
\label{table:params}
\end{table}

\subsection{Data Sets}

We use three data sets for our experiments, whose distributions are shown in
Figure~\ref{fig:datadist}. The \emph{pareto} data set contains synthetic data
generated from a Pareto distribution with $a = b = 1$. The \emph{span} data set
is a set of span durations of the distributed traces of requests that Datadog
received over a few hours. The durations are integers in units of nanoseconds,
and it includes a wide range of values (from $100$ to $1.9 \times 10^{12}$). The
\emph{power} dataset is the global active power measurements from the UCI
Individual Household Electric Power Consumption dataset \cite{UCI-POWER}.

\subsection{Sketch Size In Memory}

How much space a sketch takes up in memory will be an important consideration in
many applications. For each of the four sketches, the parameters chosen will
determine the accuracy of the sketch as well as its size. An increase in
accuracy generally requires a larger sketch. Figure~\ref{fig:sketchsize} plots
the sketch size in memory as $n$ increases.

We see that DDSketch (fast) can be up to twice the size of DDSketch, and that
HDR Histogram is significantly larger. Both GKArray and the Moments sketch are
much smaller, and the Moments sketch in particular is completely independent of
the size of the input.

If one runs DDSketch with a limit placed on the number of bins the sketch can
contain, when the maximum number of bins is reached, DDSketch starts to combine
the smallest bins together as needed, meaning that the lower quantile estimates
may not meet the relative accuracy guarantee. In our experiments this maximum
was never reached, and we have not found this to be an issue.
Figure~\ref{fig:nbins} plots the number of DDSketch bins for the \emph{pareto}
data set. The number of bins is around 900 for $n = 10^{10}$, less than half the
limit of 2048. It is also worth noting that the actual sketch size required for
the Pareto distribution is much smaller than the upper bounds we calculated in
Section~\ref{subsec:sizebounds}.

\begin{figure}[ht!]
  \centering
  \includegraphics[width=\columnwidth]{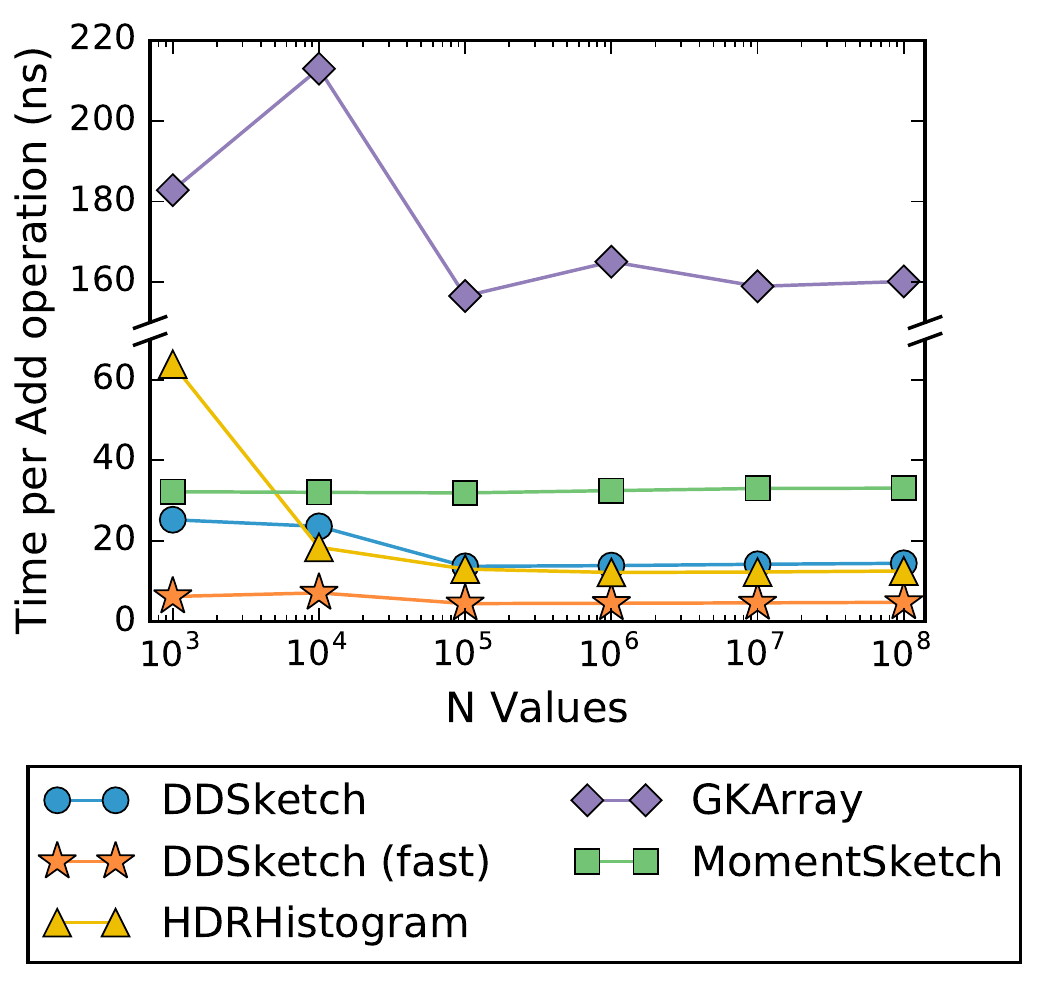}
  \caption{Average time required to add a value to a sketch.}
  \label{fig:addtime}
\end{figure}

\begin{figure}[!ht]
  \centering
  \includegraphics[width=\columnwidth]{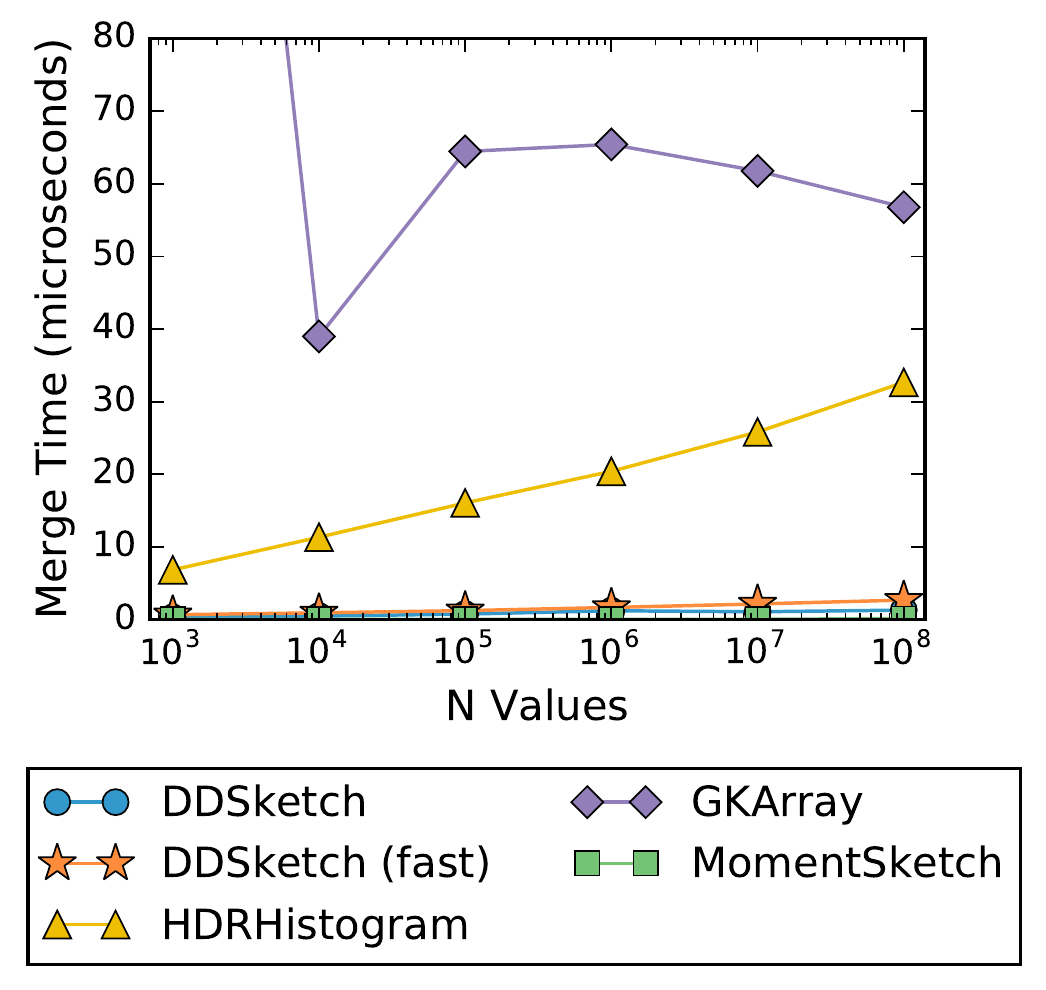}
  \caption{Average time required to merge two sketches as a function of number
    of values in the merged sketch.}
  \label{fig:mergetime}
\end{figure}

\subsection{Add and Merge Speeds}

In this section, we compare the performance of the sketches in terms of the time
required to add values to a sketch and to merge sketches
together. Figure~\ref{fig:addtime} shows the average time required to add $n$
values to an empty sketch divided by $n$. It takes less than 5 seconds to add a
hundred million values to an empty DDSketch on a 3.1GHz MacBook Pro.

GKArray is the slowest for insertions by far, being around six times slower than
the Moments sketch. Adding to an HDR Histogram is faster than adding to the
standard version of DDSketch as HDR Histogram has a simpler index calculation
than DDSketch which has to calculate logarithms. DDSketch (fast) is the fastest
in terms of insertion speed, though this was obtained by an increase in the
sketch size as we saw in Section~\ref{fig:sketchsize}.

Figure~\ref{fig:mergetime} plots the average time required to merge two sketches
of roughly the same size, as a function of the number of values in the merged
sketch. Merging two DDSketches is very fast---it takes around 10 microseconds or
less to merge two sketches containing up to fifty million values
each---depending on the data set and size, it can be more than an order of
magnitude faster than GKArray or HDR Histogram. The Moment sketch has the
fastest merge speeds of all the algorithms, as each sketch only holds on to
$k=20$ values.

\subsection{Sketch Accuracy}

\begin{figure*}[!ht]
  \centering
  \includegraphics[width=0.9\textwidth]{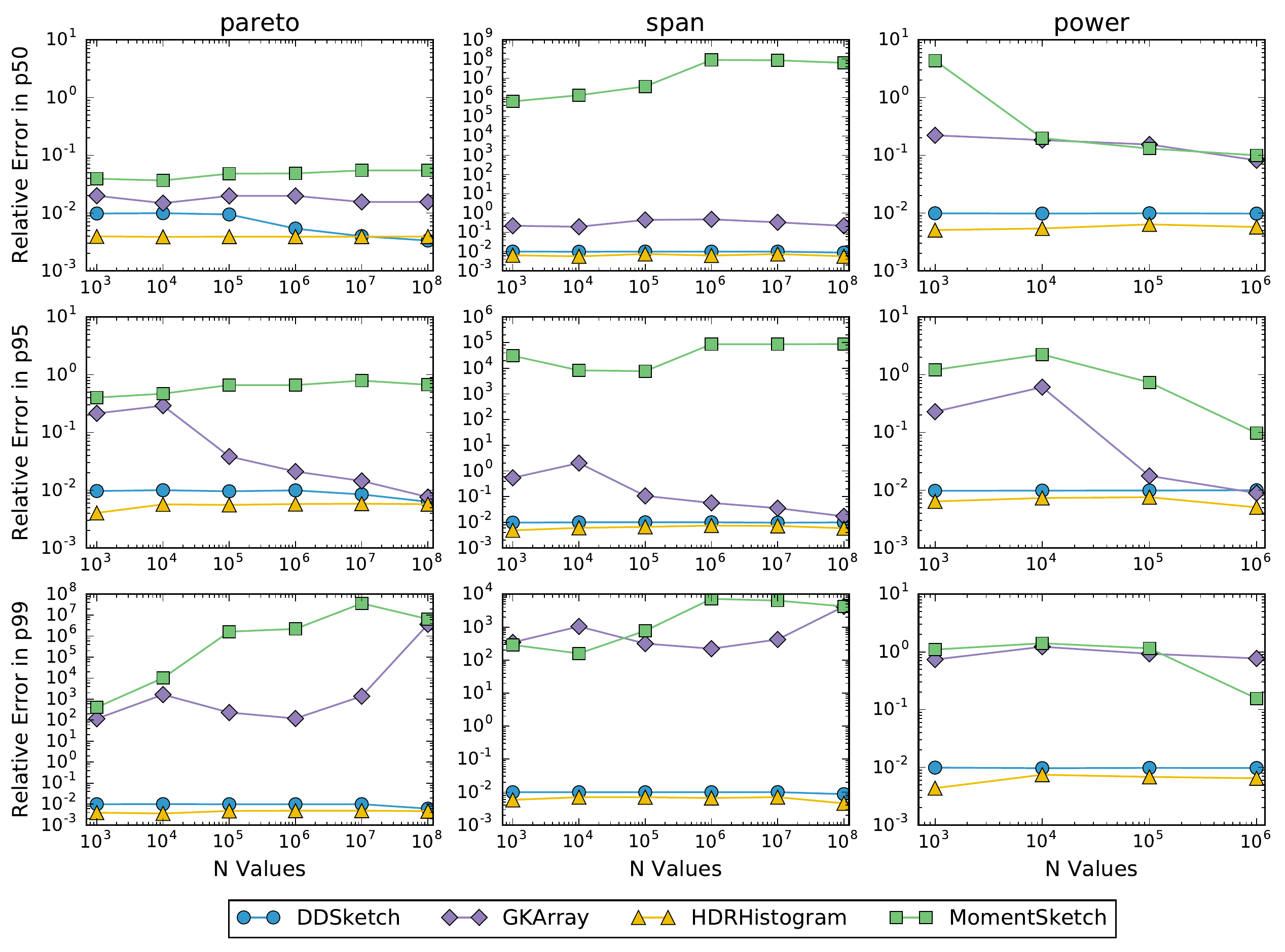}
  \caption{Relative errors of p50, p95, and p99 estimates.}
  \label{fig:relerr}
\end{figure*}

\begin{figure*}[!ht]
  \centering
  \includegraphics[width=0.9\textwidth]{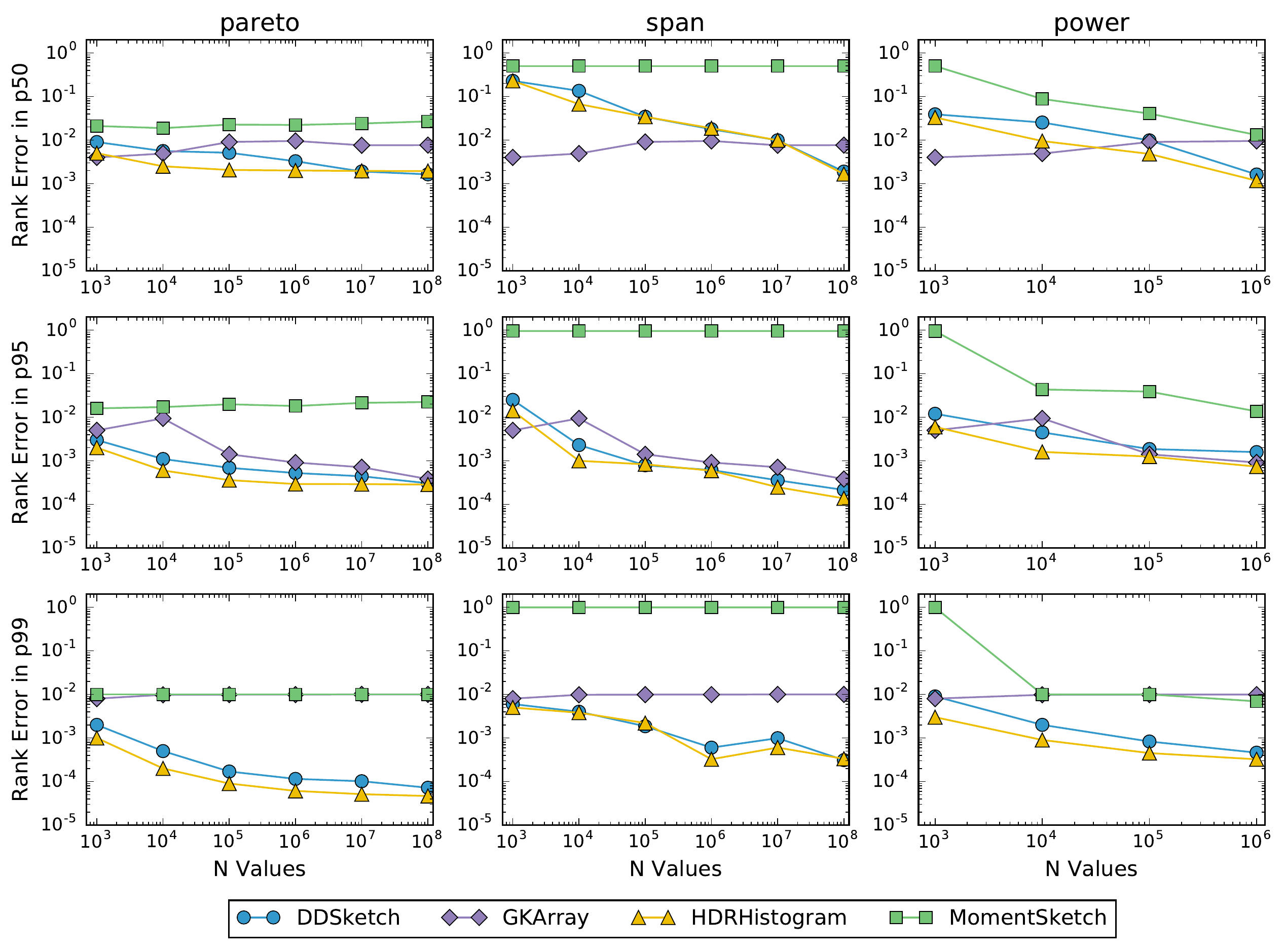}
  \caption{Rank errors of p50, p95, and p99 estimates.}
  \label{fig:rankerr}
\end{figure*}

DDSketch guarantees a relative error in its quantile estimates of at most
$\alpha$, while GKArray guarantees a rank error of less than $\epsilon$. HDR
Histogram has an implied relative-error guarantee of $10^{-d}$ where $d$ is the
number of significant digits. Therefore we compare both the average relative and
rank errors in Figures~\ref{fig:relerr} and \ref{fig:rankerr}, for the p50,
p95, and p99 estimates. Note that for GKArray, for $n \leq 1/\epsilon$, all the
values are retained so that both the relative error and rank error will be zero.

Figure~\ref{fig:relerr} shows that for all three data sets DDSketch has a
consistent relative error less than $\alpha$ for all values of $n$.  For the
heavy-tailed \emph{pareto} and \emph{span} data sets, the relative error
sketches (DDSketch and HDR Histogram) have much smaller relative error than
either GKArray or Moments. The discrepancy is especially striking for the higher
quantiles, as the values returned can be several orders of magnitude off the
actual value. The Moments sketch has particular difficulty with the \emph{span}
data set as it has trouble dealing with such a large range of values.

In terms of rank error, the guarantee of GKArray can be clearly seen in
Figure~\ref{fig:rankerr}. No such guarantee is provided for DDSketch and HDR
Histogram, yet they perform better than the Moments sketch which has a guarantee
\emph{on average}, and even GK for the higher quantiles.

\section{Conclusion}

Datadog's use-case for distributed quantile sketching comes from our agent-based
monitoring where we need to accurately aggregate data coming from disparate
sources in our high-throughput, low-latency, distributed data processing engine.
To get a sense of the scale, some of our customers have endpoints that handle
over 10M points per second, and DDSketch provides accurate latency quantiles for
these endpoints.

After our initial evaluation of existing quantile sketching algorithms,
we settled on the Greenwald-Khanna algorithm as it could handle arbitrary
values, and provided the best compromise between accuracy, size, insertion
speed, and merge time. (The implementation we provide comes from our work in
optimizing the algorithm.)

Unfortunately, the relative-accuracy errors for higher quantiles generated by
the rank-error sketch proved to be unacceptable, which led us to develop
DDSketch. Unlike HDR Histogram, which is designed to handle a bounded range and
has poor merge speeds, DDSketch is a flexible relativer error sketch that can
handle arbitrarily large ranges, and has fast merge speeds. Compared to GK, the relative
accuracy of DDSketch is comparable for dense data sets, while for heavy-tailed
data sets the improvement in accuracy can be measured in orders of magnitude.
The rank error is also comparable to if not better than that of
GK. Additionally, it is much faster in both insertion and merge.

\section{Acknowledgment}

This research has made use of data provided by the International Astronomical
Union's Minor Planet Center.

\balance

\bibliographystyle{abbrv}
\bibliography{ddsketch}
\balance

\end{document}